\documentclass[prl,showpacs,preprintnumbers,twocolumn]{revtex4}
\usepackage{amsmath}
\usepackage{amssymb}
\usepackage{color}
\usepackage{graphicx}

\renewcommand{\Re}{\mathop{\mathrm{Re}}}
\renewcommand{\Im}{\mathop{\mathrm{Im}}}

\begin{document}

\title{Scale-dependent correction to the dynamical conductivity of a disordered system at unitary symmetry}
\author{P.\ M.\ Ostrovsky}
\affiliation{Max Planck Institute for Solid State Research, Heisenbergstr.\ 1,
70569 Stuttgart, Germany}
\affiliation{L.\ D.\ Landau Institute for Theoretical Physics, 142432
Chernogolovka, Russia}
\author{Tomoyuki Nakayama and K.\ A.\ Muttalib}
\affiliation{Department of Physics, University of Florida, Gainesville FL 32611-8440}
\author{P.\ W\"olfle}
\affiliation{Institute for Condensed Matter Theory and Institute for Nanotechnology,
Karlsruhe Institute of Technology, D-76128 Karlsruhe, Germany}

\begin{abstract}
Anderson localization has been studied extensively for more than half a century.
However, while our understanding has been greatly enhanced by calculations based
on a small $\epsilon$ expansion in $d=2+\epsilon$ dimensions in the framework of
non-linear sigma models, those results can not be safely extrapolated to $d=3$.
Here we calculate the leading scale-dependent correction to the
frequency-dependent conductivity $\sigma(\omega)$ in dimensions $d \leq 3$ . At
$d=3$ we find a leading correction $\Re{\sigma(\omega)} \propto |\omega|$, which
at low frequency is much larger than the $\omega^{2}$ correction deriving from
the Drude law. We also determine the leading correction to the renormalization
group $\beta$-function in the metallic phase at $d=3$.  
\end{abstract}


\maketitle

\textit{Introduction:} Anderson localization of quantum particles in a
random potential or of classical waves in a random medium has been studied
intensively since the phenomenon was first proposed in a seminal paper by
P.\ W.\ Anderson in 1958 \cite{anderson}. In these systems localization is a
consequence of interference of multiple scattering processes, affecting the
motion of particles or waves the more, the stronger the disorder or the more
restricted the available geometry is, and leading to complete localization
of quantum particles at all energies in dimensions $d\leq 2$. In dimensions 
$d>2$ a quantum phase transition from metallic to insulating behavior takes
place as the disorder strength $\lambda$ is increased beyond a certain
threshold value $\lambda _{c}$. There are convincing arguments that the
transition is continuous and that there exists a critical point at $\lambda
=\lambda _{c}$, characterized by a diverging correlation length $\xi$ (the
localization length on the insulating side of the transition). The behavior of
physical observables such as the conductivity $\sigma$ and the dielectric
function $\epsilon$, as well as $\xi $, are described in terms of power laws in
$(\lambda -\lambda _{c})$, with critical exponents. It has been one of the
goals of the theory to determine these critical exponents.

Assuming that there is only one length scale ($\xi $) in the system,
Abrahams et al. \cite{abrahams} have proposed a scaling theory for the
dimensionless conductance $g$ as a function of the length $L$ of the sample
(considering only hypercubic systems), which takes the form of a
renormalization group equation: 
\begin{equation}
\frac{d\ln g}{d\ln L}=\beta _{L}(g). 
\label{RG}
\end{equation}
The $\beta $-function has been determined in the asymptotic regimes of very
large (metallic) and very small (insulating) $g$. On the basis of these
limiting behaviors it is plausible to conclude that a metal-insulator
transition occurs in $d=3$ dimensions at $g\thickapprox O(1)$. At these
intermediate values of $g$ a systematic and controlled calculation of the
$\beta$-function is not easily possible.

The scaling hypothesis has been justified to a large extent by a mapping of
the original problem on to an effective field theory (nonlinear $\sigma$-model)
of interacting matrices, first proposed by Wegner \cite{wegner}.
Within that model one confirms the result that all states are localized in
$d\leq 2$ dimensions. In order to determine critical exponents, the
$\beta$-function has been calculated in dimensions $d=2+\epsilon $, $\epsilon
\ll 1$, when the critical point is shifted to large values of conductance,
$g_{c}\gg 1$, and a loop expansion in powers of $1/g$ is feasible. This program
has been carried out for the three main symmetry classes: orthogonal, unitary,
and symplectic symmetry \cite{efetov}. In this way the $\beta $-function has
been found, e.g., in the orthogonal case up to terms of fourth order in $1/g$
with coefficients calculated to linear order in $\epsilon$ \cite{hikami,
wegner}.

In the unitary case the leading correction terms to the $\beta $-function
have been calculated \cite{wegner}: 
\begin{equation}
\beta _{L}(g)=\epsilon -\frac{c_{2}}{g^{2}}+O\left(g^{-5}\right).
\end{equation}
The latter result shows that in the limit of $\epsilon \rightarrow 0$ the
term linear in $1/g$ is absent. As a consequence, the critical exponent of
the conductivity turns out to be $s=1/2\epsilon$ which extrapolates to $s =
1/2$ at $d=3$, and is much too small compared to the value of $s \approx 1.3$
obtained in numerical studies \cite{slevin}. For a review of the scaling theory
see \cite{leerama}.

Here we propose an alternative strategy designed to access the $\beta $-function
in $d=3$ dimensions in the metallic regime: a direct calculation of the leading
terms in an expansion in $1/g$ in perturbation theory. Here we present first
results for the unitary case. It is well known that the leading (one-loop)
contribution is exactly zero in this case as the so called Cooperon propagator
acquires a mass, removing its diffusion pole. As a consequence, no scale
dependent ($L$-dependent) contribution to the conductance $g(L)$ is present in
the lowest order. Scale dependent terms growing with $L$ may only be generated
by infrared divergent building blocks, i.e. diffusion propagators. It may be
shown that diagrams containing one diffusion pole and any other elements may be
combined to cancel out exactly \cite{vw}. The argument carries over to higher
order terms, involving more than one diffusion pole (but no Cooperon poles, of
course), considered in the diffusive regime. This property is a consequence of
gauge symmetry.

Early attempts to calculate the quantum correction to the conductivity in
the unitary symmetry case in perturbation theory (rather than from the
nonlinear sigma model using dimensional regularization) were discouraged by
the above mentioned theorem \cite{vw}. However, by keeping track of the
contributions from larger momenta $q\gtrsim 1/l$, where $l$ is the mean free
path (ballistic regime), the correct result is recovered \cite{ostrovsky}. This
calculation manifestly obeys the gauge symmetry property discussed above and
confirms the result obtained earlier within the dimensional regularization
scheme. We now generalize the method to arbitrary dimensions $d$, in particular
$d=3$. We determine the scale-dependent contributions to the conductance in the
lowest nonzero order in $1/g$ and from there derive corrections to the
$\beta$-function in $d$ dimensions.

\textit{Dynamical conductivity:} In order to access the scale-dependent terms in
perturbation theory, from which the $\beta$-function may be derived, it is
convenient to consider the dynamical conductivity $\sigma(\omega)$ of the
infinite system as a function of positive Matsubara frequency $\omega$, rather
than the conductance $g(L)$ at $\omega=0$ as a function of system length $L$.
The conductivity of an isotropic disordered system (short range disorder) is a
function of the electron density (characterized by the Fermi wave number
$k_{F}$), and the disorder strength (characterized by the mean free path $l$)
and obeys in the scaling regime the scaling property, expressed in terms of the
correlation length $\xi$
\begin{equation}
\sigma(\omega; l,k_{F}) = \xi^{2-d}G(\omega \xi ^{z}),
\label{omega-scaling}
\end{equation}
where $z$ is the dynamical critical exponent, known to be $z=(d-2)/d$
\cite{shapabr}. We introduce the characteristic length $L_{\omega} =
(D/\omega)^{1/2}$, where $D(\omega) = \sigma(\omega)/\nu _{d}$ is the diffusion
coefficient and $\nu _{d}$ is the density of states at the Fermi level. In the
limit $\omega \rightarrow 0$ the scaling function tends to a constant, and
therefore $\xi = [\sigma (\omega =0)]^{1/(2-d)}$. The scaling law Eq.\
(\ref{omega-scaling}) holds in the neighborhood of the quantum critical point
(QCP). There are
subdominant corrections to the scaling form, the most familiar one being
generated by the Drude law (using units of electrical charge $e=1$),
\begin{equation}
\sigma_0(\omega) = \frac{n_{d}}{m}\frac{\tau}{1+\omega \tau}
=\frac{\nu_{d}D_{d}}{1+\omega \tau},
\end{equation}
with $n_{d}$, $m$ the electron density and mass, $\tau$ the momentum
relaxation time, and $D_{d}=v_{F}^{2}\tau /d$ the bare diffusion coefficient.
We define the correlation length $\xi$ in the metallic regime as $\sigma(0) =
\xi ^{2-d}$. One observes that to the leading order $\xi \sim \tau ^{1/(2-d)}$,
and then $\omega \tau \sim \omega \xi^{-zd}$, in conflict with scaling. However,
the dominant (scale-dependent) contributions to $\sigma(\omega)$ in the limit
$\omega \rightarrow 0$ will obey the scaling property, as shown below.

A renormalization group equation for $\sigma (\omega )$ as a function of 
$\omega$ is obtained by considering the dimensionless conductance
$g_{\omega}(L_{\omega })=L_{\omega }^{d-2}\sigma (\omega )$ as a function of
$L_{\omega}$
\begin{equation}
\frac{d\ln g_{\omega }}{d\ln L_{\omega }}=\beta _{\omega }(g_{\omega }) \; .
\end{equation}
In principle, the $\beta $-function may be obtained from a calculation of
the conductance in perturbation theory in the disorder. The dependence on
length $L_{\omega }$ may be represented in the form of a power series in $\ln
(L_{\omega }/L_{0})$, where $L_{0}$ is a reference length (later to be taken
equal to $l$). One thus has a double power series in the disorder parameter
$\lambda =1/(\pi k_{F}l)$ and $\ln (L_{\omega }/L_{0})$ 
\begin{equation}
g_{\omega }=g_{0}(\lambda )+g_{1}(\lambda )\ln (L_{\omega
}/L_{0})+g_{2}(\lambda )\ln ^{2}(L_{\omega }/L_{0})+ \cdots ,
\end{equation}
and 
\begin{equation}
g_{\nu }(\lambda )=\sum_{n=-1}^{\infty }a_{\nu n}\lambda ^{n} \; .
\end{equation}
The $\beta $-function may be extracted from this power series by taking the
derivative of $g_{\omega }$\ with respect to $\ln L_{\omega }$ and then
putting $L_{\omega }=L_{0}:$ 
\begin{equation}
\beta (g_{0})=\frac{1}{g_{0}}\; g_{1}(\lambda (g_{0})) \; .
\end{equation}
Here we defined $g_{\omega }(L_{0})=g_{0}$, and $\lambda (g_{0})$ is the
inverse of the function $g_{0}(\lambda )$.

\begin{figure}
\begin{center}
\includegraphics[angle=0, height=0.11\textwidth]{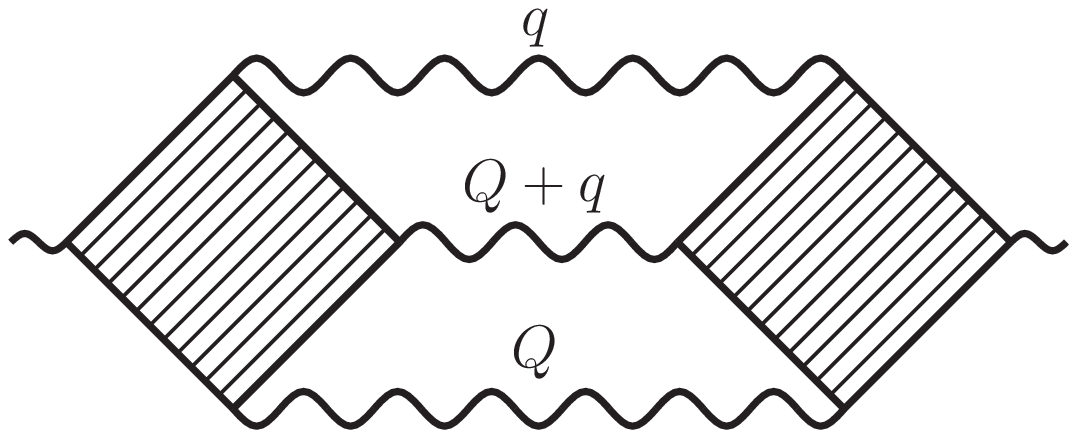}\hfill
\includegraphics[angle=0, height=0.11\textwidth]{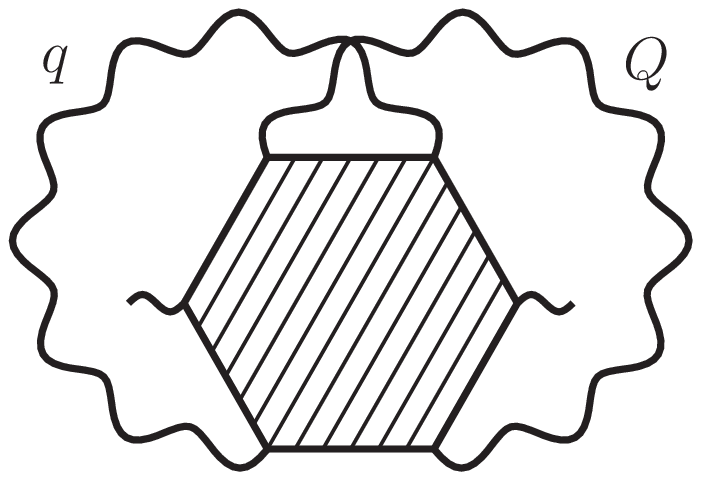}
\end{center}
\caption{Diagrams contributing to the 2-loop corrections to the conductance.
Dashed squares represent $4$-vertex vector Hikami boxes $h_{4}$, while
the hexagon represents the $6$-vertex scalar Hikami box $h_{6}$.
Wavy lines denote diffusons.}
\label{hikami-diagrams}
\end{figure}

\begin{figure}
\begin{center}
\includegraphics[angle=0, height=0.1\textwidth]{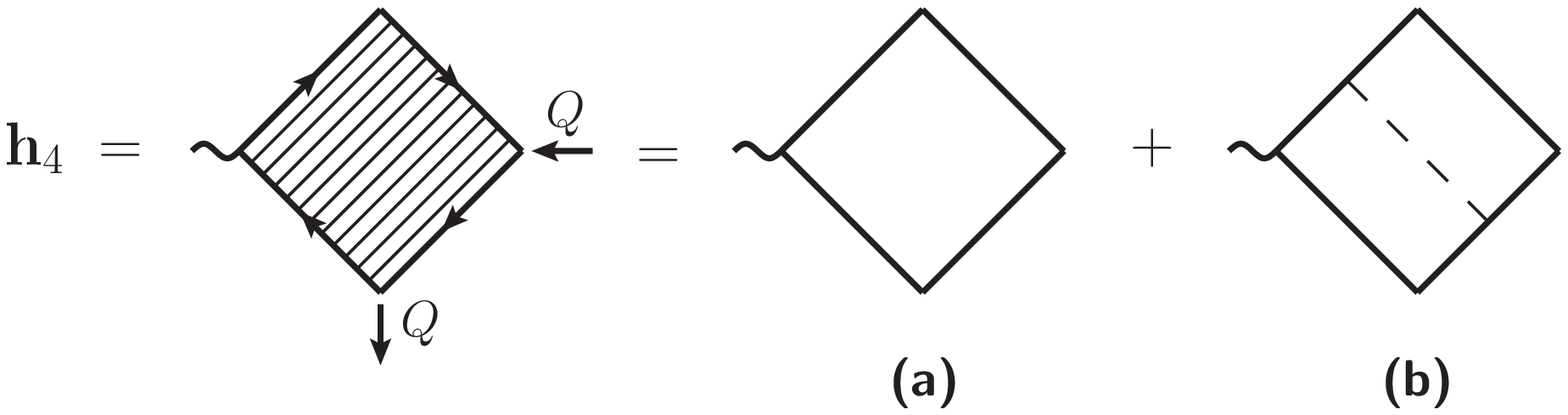}
\includegraphics[angle=0, height=0.16\textwidth]{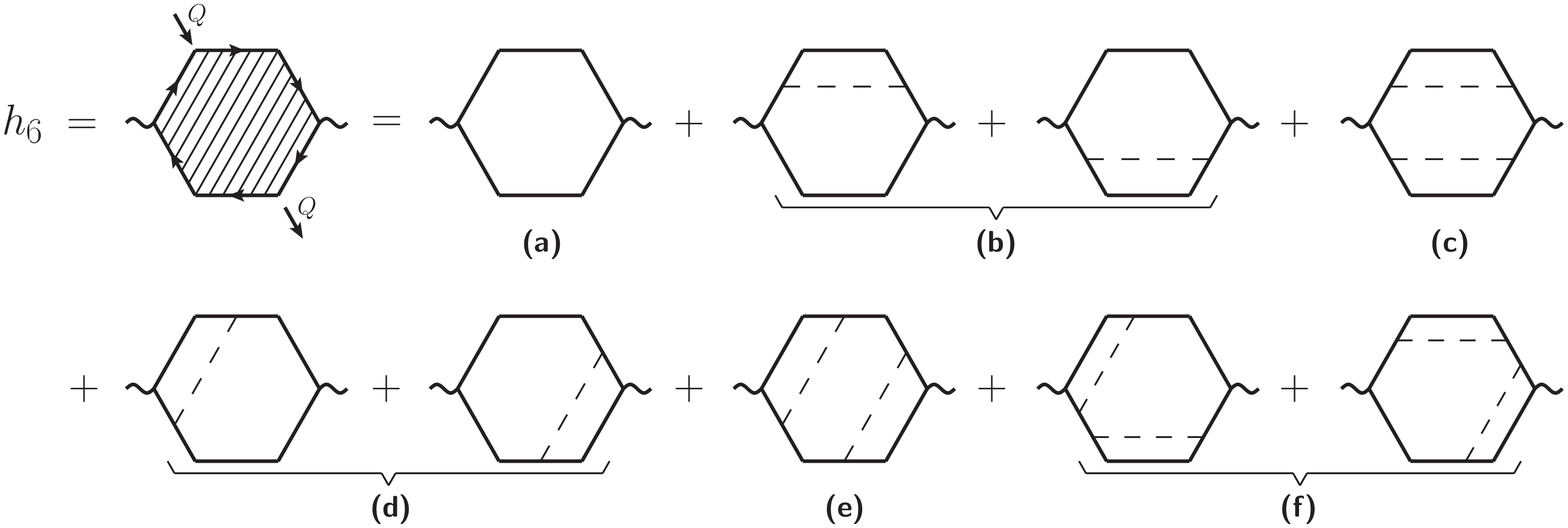}
\end{center}
\caption{Diagrams contributing to the $4$-vertex vector Hikami box $h_{4}$
(top), and the $6$-vertex scalar Hikami box $h_{6}$ (bottom). Dashed lines
represent impurity scattering.
}
\label{hikami-boxes}
\end{figure}

\begin{figure}
\begin{center}
\raisebox{11pt}{\includegraphics[angle=0, width=0.22\textwidth]{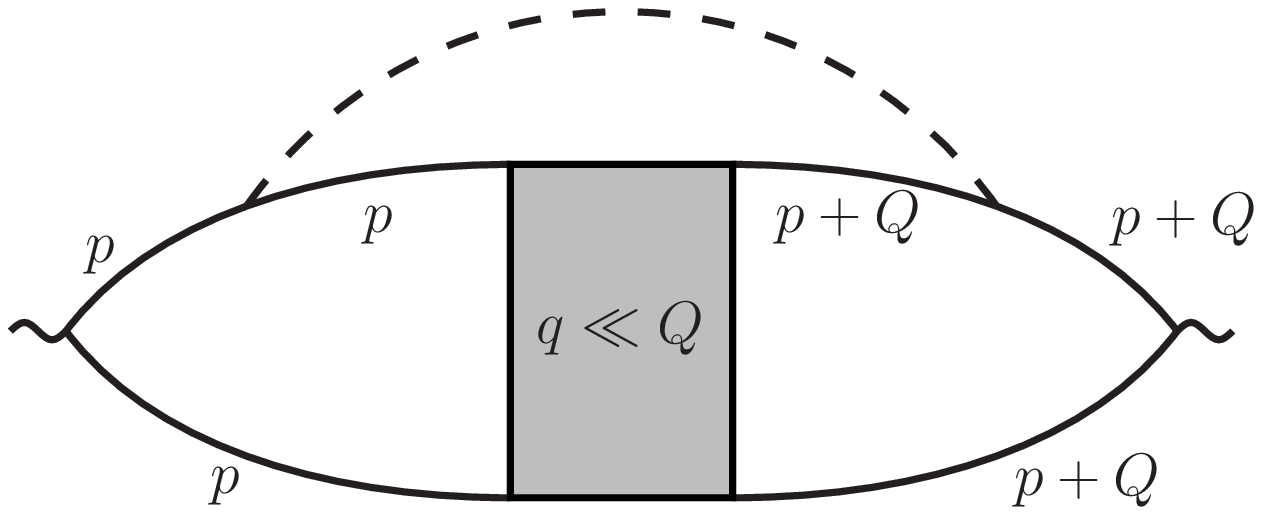}}
\includegraphics[angle=0, width=0.22\textwidth]{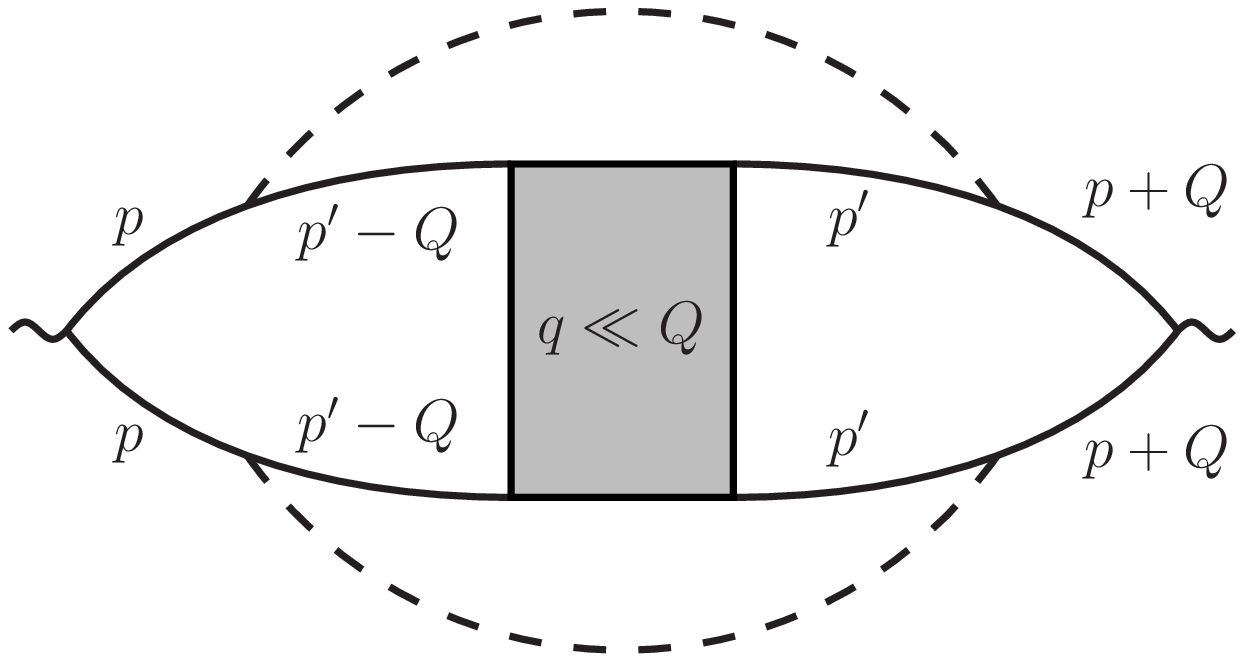}
\includegraphics[angle=0, width=0.22\textwidth]{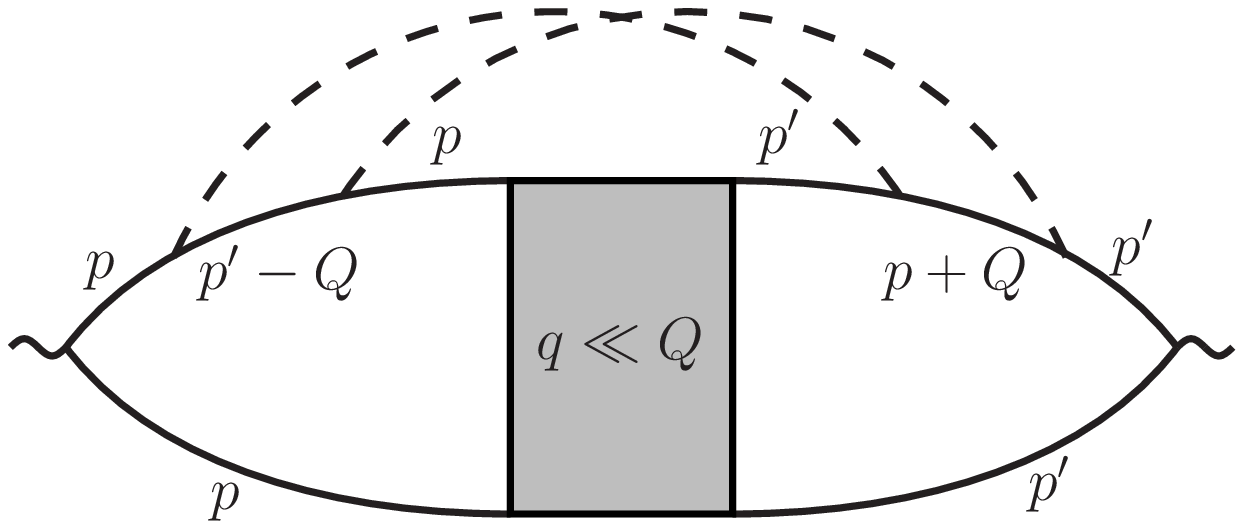}
\end{center}
\caption{The additional $2$-vertex (one-diffuson) diagrams $h_{2}$ required
by gauge invariance for 2-loop corrections to the conductance. }
\label{gaugeinv-diagram}
\end{figure}

\textit{Perturbation theory:} Scale-dependent quantum corrections to the
conductivity are generated by diagrams with infrared divergent diffusion
propagators (diffusons). As shown by Hikami \cite{hikami} the perturbation
theory may be organized in terms of diffusons connected by ``Hikami boxes''.
At two-loop order we may distinguish diagrams with two and with three
diffusons (see Fig.\ \ref{hikami-diagrams}), with $4$-vertex and $6$-vertex
Hikami boxes $h_{4}$ and $h_{6}$, respectively (see Fig.\ \ref{hikami-boxes}). 
In addition, as shown in Ref.\ \cite{ostrovsky}, gauge invariance requires a
family of one-diffuson diagrams $h_{2}$ to be added within the perturbation
scheme (see Fig.\ \ref{gaugeinv-diagram}).

In the lowest order, we then obtain the following contribution to $\sigma
(\omega)$:
\begin{align}
\sigma_\text{2-loop}(\omega)
 &= \frac{1}{2\pi}\int \frac{d^{d}q}{(2\pi )^{d}}
    D(q,\omega) \int \frac{d^{d}Q}{(2\pi)^{d}} F(\mathbf{q,Q},\omega),
\label{sigma2}\\
F(\mathbf{q,Q},\omega)
 &= D(Q,\omega) \bigg[
      \frac{1}{d}\, \mathbf{h}_{4}^{2}(q,Q) D(\mathbf{q+Q},\omega) \notag\\
 &\hspace{25mm}+ h_{6}(q,Q) + h_{2}(Q) \bigg].
\end{align}
In this expressions we have singled out one diffuson with momentum $q$ and
collected all the remaining factors in the integral of the function
$F(\mathbf{q,Q},\omega)$. 

Quite generally the diffusion propagator $D(q,\omega )$ is obtained by summing
the particle-hole ladder diagrams as 
\begin{equation}
D(q,\omega)=\frac{n_\text{imp}u_{0}^{2}}{1-n_\text{imp}u_{0}^{2}\Pi (q,\omega)},
\end{equation}
where $u_{0}$ is a short range impurity scattering potential, $n_\text{imp}$ is
the impurity density, and 
\begin{equation}
\Pi (q,\omega ) = \frac{1}{V}\sum_{\mathbf{k}}G_{\mathbf{k}+
\mathbf{q}}^{R}(E_{F}+i\omega )G_{\mathbf{k}}^{A}(E_{F})
\end{equation}
Here the retarded (advanced) Green's function of electrons of mass $m$ is given
by
\begin{equation}
G_{\mathbf{k}}^{R}(E)=\left[G_{\mathbf{k}}^{A}(E)\right]^*
=\frac{1}{E_{F}-k^{2}/2m+i/2\tau },
\end{equation}
with $E_{F}$ the Fermi energy and $\tau$ the single particle relaxation time,
defined in terms of the disorder potential by $\tau^{-1} = 2\pi \nu_{d}
n_\text{imp} u_{0}^{2}$.

We are interested in the momentum regime $q<k_{F}$, anticipating that all
momentum integrals are convergent and the contribution from the momenta
$q>k_{F}$ is negligible ($k_{F}$ is the Fermi wave vector). We consider the weak
disorder regime only, where $k_{F}l \gg 1$. In the limit of small momentum, $q l
\ll 1$ the diffuson has the usual diffusion pole form
\begin{equation}
D(q,\omega ) = \frac{1}{2\pi \nu _{d}\tau ^{2}} \frac{1}{D_{d}q^{2}+\omega}.
\label{diffpole}
\end{equation}
In the same diffusive limit $ql,Ql \ll 1$, one has to retain only the diagram
(a) in the 4-vertex Hikami box $h_4$ and the diagrams (a-c) for the 6-vertex
Hikami box $h_6$ (see Fig.\ 2) while the single diffuson diagrams of Fig.\ 3
are not important. This yields
\begin{align}
\mathbf{h}_{4}^{2}(\mathbf{Q},\mathbf{q}) &=-(4\pi \nu
_{d}D_{d})^{2}\tau^{6}(\mathbf{Q+q})^{2}, \\
h_{6}(Q,q) &= 4\pi \nu _{d}D_{d}\tau ^{4}, \\
h_{2}(Q) &=0.
\end{align}

We now substitute these results into the above expression for the
conductivity, first concentrating on the contribution from the diffusive
regime
\begin{multline}
\sigma_\text{2-loop}^{d}(\omega) = \frac{D_{d}}{2\pi^{2}\nu_{d}}\int
\frac{d^{d}q\,d^{d}Q}{(2\pi)^{2d}} \frac{1}{D_{d}q^{2}+\omega
}\frac{1}{D_{d}Q^{2}+\omega }\\
\times \left[1-\frac{2D_{d}(\mathbf{q+Q})^{2}/d}{D_{d}(\mathbf{q+Q})^{2}+\omega
}\right].
\label{sigma-2loop}
\end{multline}
In low dimensions, $d<2$, the integrals are convergent, yielding the
scale-dependent result
\begin{multline}
\sigma_\text{2-loop}^{d}(\omega) = C_{d<2}\nu_{d}D_{d}\left[\omega
\nu_{d}^{2/(2-d)}D_{d}^{d/(2-d)}\right]^{d-2} \\
= C_{d<2}\xi^{2-d}\left[\frac{L_{\omega}}{\xi }\right]^{4-2d},
\label{Eq-Cdless}
\end{multline}
where the constant is given by 
\begin{multline}
C_{d<2} = \frac{1}{d\pi ^{2}}\int \frac{d^{d}x\,d^{d}y}{(2\pi)^{2d}}
\frac{1}{x^{2}+1}\frac{1}{y^{2}+1}\\
 \times  \left[\frac{d}{2}-1+\frac{1}{(\mathbf{x+y})^{2}+1}\right].
\end{multline}
The integrals may be reduced to a single integral of a hypergeometric
function
\begin{multline}
C_{d<2} =  2^{-2d}\pi ^{-d-2}\Bigg[\!\left(\frac{1}{2}-\frac{1}{d}\right)
\Gamma^{2}\left(1-\frac{d}{2}\right) + \frac{2\Gamma (d-3)}{d(4-d)} \\
\times
\int_{0}^{1}dt\;F\left(2-\frac{d}{2},\frac{d}{2},3-\frac{d}{2},1-t+t^{2}
\right)\Bigg].
\label{Eq-CdH}
\end{multline}
with $C_{1}=-1/(24\pi ^{2})$ and $C_{2-\epsilon }=-1/(16\pi ^{4}\epsilon)$,
at $\epsilon \ll1$. Thus, in the limit $\epsilon \to 0$, we find the
logarithmic scaling dependence
\begin{multline}
\sigma_\text{2-loop}^{2-\epsilon}(\omega ) = -\frac{1}{16\pi ^{4}\nu_{2}D_{2}}
\frac{\omega ^{-\epsilon }}{\epsilon } \\ 
\rightarrow \frac{\ln \omega}{16\pi ^{4}\nu_{2}D_{2}}
=-\frac{\ln L_{\omega }}{8\pi ^{4}\nu _{2}D_{2}}, \label{sigma_d=2}
\end{multline}
which is in agreement with the known two-loop correction result to the
conductivity in $d=2$ dimensions (here obtained by approaching $d=2$ from
below).

The same result can be obtained directly in $d = 2$, without resorting to the
$\epsilon$ expansion. If we assume that $q < Q$, then the infrared-divergent
part of Eq.\ (\ref{sigma2}) originates from the integral over $q$, while
the $Q$-integral yields the numeric prefactor. The diffusons and Hikami boxes
entering the function $F(\mathbf{q,Q},\omega)$ should be calculated with an
increased precision allowing for high (ballistic) values of $Q$. This requires
taking into account all the diagrams shown in Figs.\ 2 and 3. Ballistic
contributions make the integral over $Q$ convergent leading to the same
logarithmic correction (\ref{sigma_d=2}).

We now turn to dimensions $d>2$, for which the momentum integrals in
Eq.\ (\ref{sigma-2loop}) are not converging. In order to regularize these
integrals we first take a logarithmic derivative with respect to frequency
$\omega$ 
\begin{equation}
\frac{d}{d\ln \omega}\sigma_\text{2-loop}^{d}(\omega ) = I_{1}+I_{2}
\end{equation}
where
\begin{align}
I_{1} &=-\frac{D_{d}}{\pi ^{2}\nu _{d}} \int \frac{d^{d}q\,d^{d}Q}{(2\pi)^{2d}}
\frac{\omega}{(D_{d}q^{2}+\omega )^{2}}\frac{1}{D_{d}Q^{2}+\omega }\\
&\hspace{3cm} \times \left[
1-\frac{2D_{d}(\mathbf{q+Q})^{2}/d}{D_{d}(\mathbf{q+Q})^{2}+\omega}
\right], \\
I_{2} &=\frac{D_{d}}{d\pi ^{2}\nu_{d}} \int \frac{d^{d}q\,d^{d}Q}{(2\pi)^{2d}}
\frac{\omega}{D_{d}q^{2}+\omega}\\
&\hspace{21mm}\times \frac{1}{D_{d}Q^{2}+\omega}
\frac{D_{d}(\mathbf{q+Q})^{2}}{[D_{d}(\mathbf{q+Q})^{2}+\omega]^{2}}.
\end{align}
While $I_{2}$ converges provided $d<3$, (for a further regularization see
below), $I_{1}$ is still ultraviolet divergent. Including the contribution
to the momentum integral on $Q$ from the ballistic regime we have  
\begin{multline}
I_{1} = -\frac{1}{4\pi ^{2}\nu _{d}\tau ^{2}}\int \frac{d^{d}q}{(2\pi )^{d}}
\frac{\omega }{(D_{d}q^{2}+\omega )^{2}}\\
\times \int \frac{d^{d}Q}{(2\pi )^{d}}F(\mathbf{q,Q},\;\omega).
\label{Eq-I1}
\end{multline}
We now use the identity $\int d^{d}QF(0, \mathbf{Q}, 0)=0$, which may be checked
by direct calculation and, more generally, is a consequence of the gauge
invariance \cite{vw}. Therefore, replacing $F(\mathbf{q,Q}, \omega )$ by 
$F(\mathbf{ q,Q}, \omega) - F(0, \mathbf{Q},0)$ in Eq.\ (\ref{Eq-I1}), we obtain
convergent integrals in the ultraviolet, which allows us to use the diffusive
limit expressions for $F(\mathbf{q,Q}, \omega)$. Adding the contribution
$I_{2}$, and reformulating the integrals using the symmetry under
$\mathbf{q\leftrightarrow Q}$, we then find
\begin{multline}
\frac{d}{d\ln \omega}\sigma_\text{2-loop}^{d}(\omega)
=\frac{d}{d\ln\omega} \frac{D_{d}}{d\pi ^{2}\nu _{d}}\\
\times \int \frac{d^{d}q\,d^{d}Q}{(2\pi)^{2d}}
\frac{\omega}{D_{d}q^{2}+\omega }\frac{1}{D_{d}Q^{2}+\omega } \\
\times \left[\frac{1}{D_{d}(\mathbf{q+Q})^{2}+\omega}
-\frac{(1-d/2)\omega}{D_{d}^{2}q^{2}Q^{2 }}\right].
\end{multline}
The scaling contribution to the conductivity in dimensions $2 < d < 3$ takes the
same form as in Eq.\ (\ref{Eq-Cdless}) where the constant $C$ is now given by
\begin{multline}
C_{d>2} = \frac{1}{d\pi ^{2}}\int
\frac{d^{d}x\,d^{d}y}{(2\pi)^{2d}}\frac{1}{x^{2}+1}\frac{1}{y^{2}+1}\\
\times 
\left[\frac{\frac{d}{2}-1}{x^{2}y^{2}}+\frac{1}{(\mathbf{x+y})^{2}+1}\right]\; .
\end{multline}
The integrals may be done to give the same expression as for $d<2$, namely
Eq.\ (\ref{Eq-CdH}). This means that by analytic continuation in the complex
$d$-plane one
may pass from the regime $d<2$ to the regime $d>2$, around the singularity
at $d=2$. For dimensions close to integer values we get $C_{2+\epsilon}
= 1/(16\pi ^{4}\epsilon) = -C_{2-\epsilon}$, and $C_{3-\epsilon } =
1/(96\pi^{4}\epsilon)$. 

The logarithmic ultraviolet divergence in the case of three dimensions is
cutoff by the upper limit $q=q_{0}\approx 1/l$, leading to a logarithmic
contribution
\begin{equation}
\sigma_\text{2-loop}^{d=3}(\omega )
 = -\frac{\omega \ln \omega}{96\pi ^{4}\nu_{3}D_{3}^{2}}
 \rightarrow \frac{1}{48\pi ^{4}} \xi^{-1}
   \frac{\ln (c L_{\omega}/\xi)}{(L_{\omega }/\xi)^{2}}.
 \label{sigma3scale}
\end{equation}
Subleading corrections requiring a precise calculation of the
contribution from the ballistic regime lead to the factor $1/\xi$ and the
constant $c$ under the logarithm in the last equation. We now analytically
continue from the imaginary frequency axis to the real axis by replacing the
Matsubara frequency $\omega$ by the real frequency $\Omega=i\omega$. We note
that 
$\Re \sigma_\text{2-loop}^{d=3}(\Omega) \sim |\Omega|$
and 
$\Im \sigma_\text{2-loop}^{d=3}(\Omega) \sim \Omega \ln |\Omega|$
in that case. 

It is remarkable that the correction term to $\Re \sigma(\Omega)$ varies
linearly with $\Omega$,
\begin{equation}
\Re\sigma_\text{2-loop}^{d=3}(\Omega) = \sigma_{0} \frac{9\pi}{32} (k_{F}l)^{-3}
\frac{|\Omega|}{\epsilon_{F}}
\label{sigma3}
\end{equation}
and is thus dominant at frequencies $\Omega < \Omega_{s}$, compared to the
leading finite frequency correction $\sim -\sigma_{0} (\Omega \tau)^{2}$ of the
Drude law. Moreover the scale-dependent term is positive in contrast to the
Drude correction, so that we predict that the conductivity has a maximum at 
$\omega=\Omega_{s}/2$.  We estimate the crossover frequency as
\begin{equation}
\Omega_{s}=\frac{9\pi}{8} (k_{F}l)^{-5} \epsilon_{F}.
\end{equation}

At finite temperature the RG-flow is cut off by phase relaxation effects. This
is because in the presence of interaction the number of particles at given
energy is no longer conserved, giving rise to a phase relaxation term
$1/\tau_{\phi}$  in the independent particle diffusion pole expression Eq.\
(\ref{diffpole}). The phase relaxation rate is temperature dependent and
vanishes at $T=0$ (for calculations of $1/\tau_{\phi}$ in two dimensions see 
\cite{polyakov,narozhny}). At finite $T$ one may replace $\Omega$ by $\Omega +
i/\tau_{\phi}$ in  Eq.\ (\ref{sigma3scale}). At $\Omega \ll  i/\tau_{\phi}$ one
then finds $\sigma \sim \Omega^{2} \tau_{\phi}$, which is distinguished from the
Drude contribution by a temperature dependent prefactor growing with decreasing
temperature and is larger than the Drude term if $1/\tau_{\phi} < \Omega_{s}$. 

\textit{Renormalization group equation:} We are now in a position to determine
the leading term in a $1/g_{\omega }$-expansion of the RG-$\beta$-function for
any dimension $d$. For dimensions $2<d<3$, we then find, following the steps
outlined in the introduction
\begin{equation}
\frac{d\ln g_{\omega }}{d\ln L_{\omega }}=\beta _{\omega }(g_{\omega
})=d-2-2(d-2)C_{d}\frac{1}{g_{\omega }^{2}}+ \cdots .
\end{equation}
In the limit $d\rightarrow 2$, putting $d=2+\epsilon $, we recover the
known result (now approaching $d=2$ from above)
\begin{equation}
\beta _{\omega }(g_{\omega })=\epsilon -\frac{1}{8\pi
^{4}}\frac{1}{g_{\omega}^{2}}+ \cdots .
\end{equation}

The case $d=3$ requires special attention. As discussed above, in that case
a logarithmic correction factor in the scale dependence is found \cite{tom}.
Consequently, the $\beta $-function also acquires a logarithmic correction
factor
\begin{equation}
\beta _{\omega }(g_{\omega })=1-\frac{1}{48\pi ^{4}}\frac{\ln g_{\omega
}+c_{3}}{g_{\omega }^{2}}+ \cdots .
\end{equation}
The small prefactor $1/48\pi ^{4}$ makes the correction term small and
probably insignificant in the neighborhood of the transition point. The
latter statement anticipates that the constant $c_{3}$ is of order unity and
therefore does not  shift the critical point much. Preliminary calculations
of the contributions to $c_{3}$ appear to confirm this assumption. It
therefore appears that our result does not allow a reasonable determination
of the critical exponent. It rather suggests that the $\beta $-function has
a more complex shape caused by higher loop contributions.

\textit{Conclusion:} In the above we presented a derivation of the leading
scale-dependent contribution to the dynamical conductivity of a disordered
system in the unitary symmetry class within the model of non-interacting
fermions. We showed how the scaling terms may be extracted from the
perturbation theory in any dimension $d$. On the basis of these results we
determined the leading term in the renormalization group $\beta $-function in
the regime of dimensions $2\leq d\leq 3$. It is interesting
to note that the leading correction term in $\beta$ at large conductance $g$
varies as $g^{-2}$, independent of dimension, with a logarithmic correction
at $d=3$. Its prefactor is rather small, suggesting that higher loop order
terms will be important near the transition point ($\beta (g_{c})=0$). It
will therefore not be easily possible to determine the critical exponent in
that way. 

\textit{Acknowledgements:} We acknowledge illuminating discussions with E.\ 
Abrahams, I.\ V.\ Gornyi, A.\ D.\ Mirlin, V.\ E.\ Kravtsov, D.\ G.\ Polyakov,
and D.\ Vollhardt. This work has been partially supported by the DFG-Center for
Functional Nanostructures at KIT (KAM, PW).

\end{document}